\documentstyle[prl,aps,epsfig]{revtex}
\begin{document}
\title{Tight-binding study of interface states in semiconductor heterojunctions
}

\author{A.\ V.\ Kolesnikov$^{1}$, R.\ Lipperheide$^{2}$, and U.\ Wille$^{2}$}

\address{$^{1}$Fakult\"{a}t f\"{u}r Physik und Astronomie, Ruhr-Universit\"{a}t
Bochum, \\ Universit\"{a}tsstra\ss{}e 150, D-44780 Bochum, Germany\\
$^{2}$Abteilung Theoretische Physik, Hahn-Meitner-Institut Berlin,\\
Glienicker Stra\ss{}e 100, D-14109 Berlin, Germany}

\date{\today}
\maketitle

\begin{abstract}
Localized interface states in abrupt semiconductor heterojunctions are studied
within a tight-binding model.  The intention is to provide a microscopic
foundation for the results of similar studies which were based on the two-band
model within the envelope function approximation.  In a two-dimensional
description, the tight-binding Hamiltonian is constructed such that the
Dirac-like bulk spectrum of the two-band model is recovered in the continuum
limit.  Localized states in heterojunctions are shown to occur under conditions
equivalent to those of the two-band model.  In particular, shallow interface
states are identified in non-inverted junctions with intersecting bulk
dispersion curves.  As a specific example, the GaSb-AlSb heterojunction is
considered.  The matching conditions of the envelope function approximation are
analyzed within the tight-binding description.  \\[0.3cm]

\noindent
PACS number(s): 73.20.-r, 73.40.Lq, 73.20.Fz
\end{abstract}

\section{Introduction}

The advent of highly developed methods of nanotechnology has opened rich
possibilities to study quantum effects in semiconductor heterostructures, such
as heterojunctions, quantum wells, and superlattices (for a recent survey, see
Ref.\ \cite{and98}).  Continued interest is placed on the occurrence and
properties of localized electronic states in heterostructures, i.e., states
that are bound in the growth direction of the structure.  This interest arises
mainly from the effect that this kind of state is expected to have on transport
properties of nanodevices \cite{bas96,fer97}.  Basically, two different types
of localized states can be distinguished.  In quantum wells and superlattices,
such states occur owing to the presence of minima in the band edge potential
that confine the electronic motion in the growth direction (``size
quantization'').  On the other hand, the existence of localized states at the
interface of a {\em single} heterojunction, when no confining potential is
present, hinges on the coupling between the electronic motion perpendicular to
the interface plane and the free motion parallel to this plane
\cite{kol98,kol99}.  Owing to this coupling, energy can be exchanged between
perpendicular and parallel degrees of freedom.

While localized states in confining potentials have been widely studied
\cite{bas96,fer97,ivc97}, only few theoretical investigations have dealt with
localized states at single heterojunction interfaces
\cite{kol98,vol85,kor87,pan87,aga88,vol95}.  These were carried out within the
frame of the two-band model \cite{kel63,wol64} in the envelope function
approximation, using an effective Dirac-type Hamiltonian where the interaction
between the motion perpendicular and parallel to the interface is mediated by
an effective spin-orbit coupling.  It was found that in abrupt junctions with
band inversion localized interface states occur with a linear ``in-plane''
dispersion \cite{vol85,kor87}.  In a more general treatment \cite{kol98},
interface states with a nonlinear in-plane dispersion were identified in
junctions {\em without} band inversion, provided the ratio of energy gap and
effective mass changes across the interface and the bulk dispersion curves of
the two semiconductors intersect.  Recently, the two-band model has been
reformulated in terms of a one-band (Schr\"{o}dinger-like) description
\cite{kol99}.  The qualitative features for abrupt junctions were found to
persist essentially for graded junctions \cite{kol98,vol85,pan87,aga88,vol95}.

The two-band model in conjunction with the envelope function approximation
allows a simple and transparent description of localized interface states.
However, a more fundamental approach would have to take explicitly into account
the atomistic structure of the semiconductors making up the heterojunction.
Here, the tight-binding approximation, which has been widely used for the
description of bulk matter \cite{har89,har99} and surface states \cite{dav92},
offers an appropriate starting point.  In the present paper, we apply the
tight-binding method to study interface states in heterojunctions, with the aim
to provide a microscopic foundation for the results obtained from the two-band
treatment.  We formulate a tight-binding model for heterojunctions that is
based on a simplified description of III-V bulk semiconductors.  A
two-dimensional Hamiltonian with nearest-neighbor interactions is set up from
which the Dirac-like bulk spectrum of the two-band model is recovered in the
continuum limit and which allows an analytic derivation of the conditions for
the existence of interface states in heterojunctions.

The paper is organized as follows.  In Sec.\ II, the tight-binding model for
heterojunctions is formulated and a relation for the in-plane energy dispersion
of localized interface states is derived.  The correspondence between the
parameters of the two-band model and of the tight-binding model is revealed.
In Sec.\ III, the properties of the energy spectrum of interface states are
analyzed and compared to those of the two-band model; the case of the GaSb-AlSb
heterojunctions is discussed as a specific example.  The matching conditions
for the envelope functions of the two-band model are elucidated in the
tight-binding model.  Section IV contains a summary and our conclusions.

\section{Tight-binding model for heterojunctions}

\subsection{Qualitative picture}

We consider heterojunctions composed of two different materials of the III-V
class, thereby covering most of the cases of practical interest \cite{rho93}.
A reasonable description of the band structure of homogeneous III-V
semiconductors can be achieved within the tight-binding approach \cite{har89},
assuming a minimal basis set of one s-orbital and three p-orbitals on each atom
and including only nearest-neighbor interactions.  As we are looking here for
localized states lying in the fundamental energy gap of the heterojunction
(i.e., for states lying between the maximum of the valence band edge profile
and the minimum of the conduction band edge profile of the junction), we are
interested only in an accurate description of the bands of the homogeneous
semiconductors closest to their respective gap.  Then, following Ref.\
\cite{sai78} where a tight-binding calculation of the band structure of
the InAs-GaSb superlattice was performed, we can reduce the size of the basis
set by discarding the p-orbitals on the metallic atoms and the s-orbitals on
the nonmetallic atoms.

\begin{figure}
\epsfysize=9.0cm
\epsfbox[-240 319 458 800]{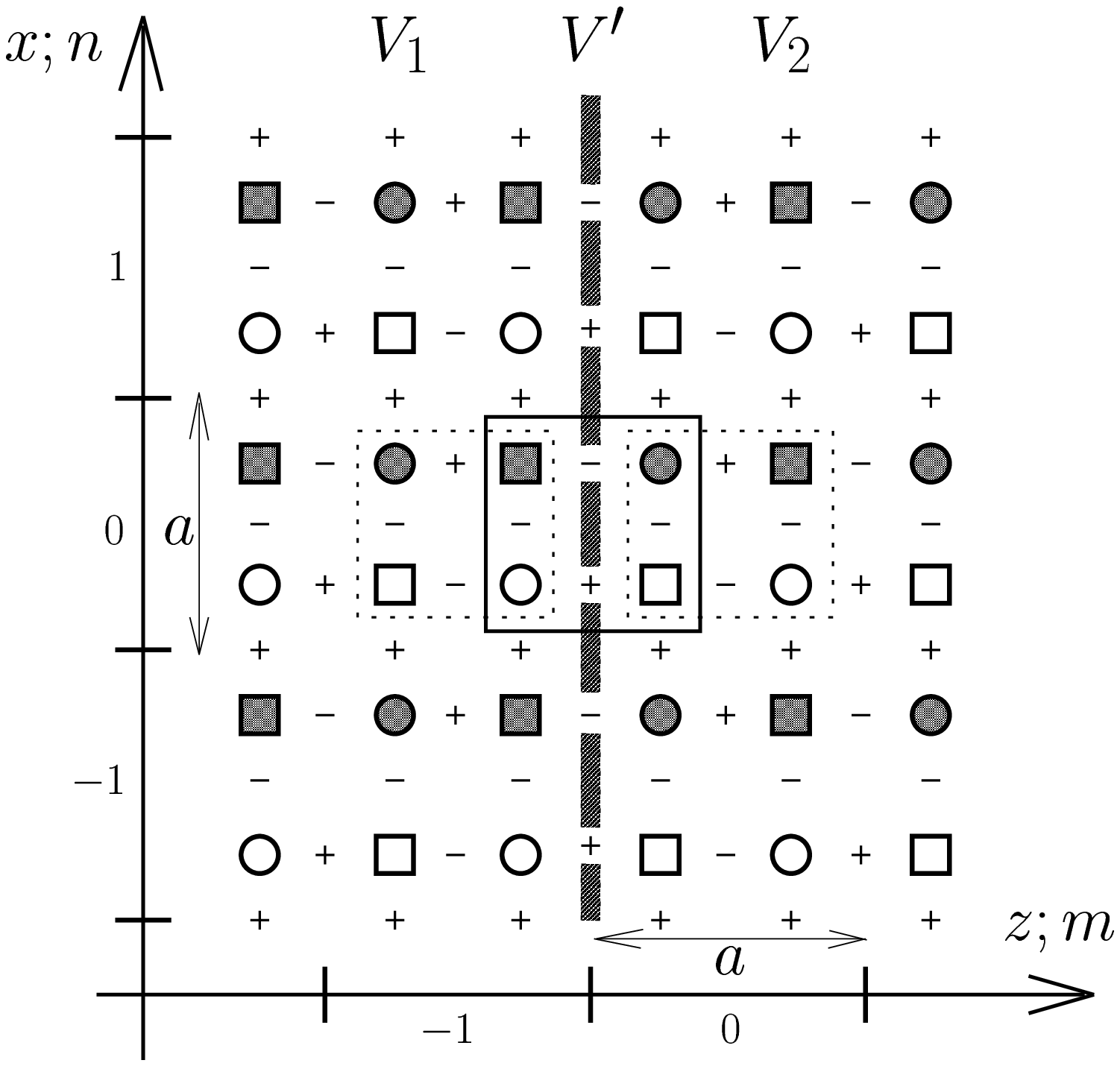}
\caption
{Two-dimensional square lattice representing a heterojunction formed by two
III-V semiconductors.  Integers attached to the $x$-axis and $z$-axis
correspond to labels $n$ and $m$, respectively.  Boxes with dashed lines
enclose unit cells with $n,m = 0,-1$ and $n,m = 0,0$, respectively [cf.\ Eq.\
(\ref{eq:15})]; the box with full lines encloses atomic sites entering Eq.\
(\ref{eq:16}).  Symbols in unit cells designate s-orbitals corresponding to
coefficients $u_{nm}$ (open boxes) and $\bar{u}_{nm}$ (shaded boxes), and
p-orbitals corresponding to coefficients $v_{nm}$ (open circles) and
$\bar{v}_{nm}$ (shaded circles) [cf.\ Eq.\ (\ref{eq:2})].  On-site energies are
$\epsilon_{\rm s}^{(j)}$ and $\epsilon_{\rm p}^{(j)}$; $j=1$ for $m < 0$ $(z <
0)$, $j=2$ for $m \geq 0$ $(z > 0)$, and the hopping matrix elements between
nearest neighbors are $\pm V_{j}$.  Hopping across the interface at $z=0$ is
described by matrix elements $\pm V'$.
}
\vspace{0.4cm}
\label{fig:1}
\end{figure}

For the actual treatment of the heterojunction, we consider a simplified,
two-dimensional model system that allows a transparent description in which the
properties of interface states can be derived essentially in analytic form.
The crucial feature of this model system is the explicit inclusion of the
coupling between the electronic motion perpendicular and parallel to the
interface plane.  Basically, a two-dimensional description appears adequate for
the present problem as in the continuum limit one deals, due to axial symmetry
about the direction perpendicular to the interface, with a truly
two-dimensional problem.

As depicted schematically in Fig.\ \ref{fig:1}, we consider an (infinitely
extended) square lattice with lattice constant $a$, where the $z$-axis is
chosen perpendicular to the interface plane (with the interface at $z=0)$ and
the $x$-axis parallel to it.  With s- and p-orbitals assigned alternatingly to
the atomic sites, the unit cells of the model system are each formed by two
s-atoms and two p-atoms.  Integers $n$ and $m$ corresponding to the
$x$-direction and the $z$-direction, respectively, are used to label the
position of the unit cells on the lattice.  The homogeneous semiconducting
materials 1 and 2 forming the heterojunction are characterized by atomic
energies $\epsilon^{(j)}_{\rm s}$ and $\epsilon^{(j)}_{\rm p}$ $(j=1,2)$, and
by the magnitude $V_{j}$ of the hopping matrix elements between
nearest-neighbor atoms, which is assumed to be independent of the hopping
direction.  As the p-orbitals are odd, the matrix elements connecting a
specific atomic orbital to its nearest neighbors in either of the two
directions differ in sign.  The interface separating the semiconductors is
located between the vertical rows of units cells with labels $m=-1$ and $m=0$.
A value $V'$ is assumed for the magnitude of the matrix elements connecting s-
and p-orbitals across the interface; in the course of the formulation of the
model, $V'$ will be expressed in terms of the bulk matrix elements $V_{1}$ and
$V_{2}$.

In the model system illustrated by Fig.\ \ref{fig:1}, the relative signs of the
hopping matrix elements in adjacent linear chains of atoms have been chosen
such that the matrix elements connecting the atoms within a unit cell add up to
a nonzero value.  By reversing in Fig.\ \ref{fig:1} the signs of all matrix
elements within every second chain either in the $x$- or the $z$-direction, one
would obtain a topologically different system in which the sum of the matrix
elements within a unit cell is zero.  The distinctive feature of the model
system of Fig.\ \ref{fig:1} is that in the continuum limit it exhibits a
Dirac-like bulk dispersion (cf.\ below).  It is noted that there is a close
analogy between the present lattice model and certain two-dimensional models
describing ``flux phases'' in strongly correlated electron systems, like
high-$T_{\rm c}$ superconductors~\cite{aff88,zha90,ubb92,fra93}.

\subsection{Homogeneous semiconductors}

\begin{figure}
\epsfysize=5.0cm
\epsfbox[-265 400 509 708]{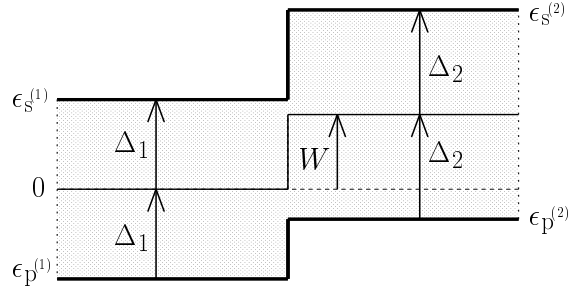}
\caption{Schematic band edge profile for an abrupt heterojunction.}
\vspace{0.4cm}
\label{fig:2}
\end{figure}

Turning now to the detailed formulation of our model, we write the
two-dimensional tight-binding Hamiltonian for a {\em homogeneous} semiconductor
characterized by atomic energies $\epsilon_{\rm s}$ and $\epsilon_{\rm p}$ and
by the magnitude $V$ of the nearest-neighbor hopping matrix elements as
\begin{eqnarray}
H &=& \sum_{n,m} \{ \epsilon^{}_{\rm s} (a^{\dagger}_{nm} a^{}_{nm} +
\bar{a}^{\dagger}_{nm} \bar{a}^{}_{nm}) + \epsilon^{}_{\rm p} (b^{\dagger}_{nm}
b^{}_{nm} + \bar{b}^{\dagger}_{nm} \bar{b}^{}_{nm})  \nonumber
\\[-0.30cm] &\ & \hspace*{0.6cm} + \; V [ (b^{\dagger}_{n,m-1} +
\bar{b}^{\dagger}_{n-1,m} - b^{\dagger}_{nm} - \bar{b}^{\dagger}_{nm} )
a^{}_{nm} + (b^{\dagger}_{n+1,m} - \bar{b}^{\dagger}_{n,m+1} +
\bar{b}^{\dagger}_{nm} - b^{\dagger}_{nm} ) \bar{a}^{}_{nm}  \nonumber
\\ &\ & \hspace*{1.3cm} + (a^{\dagger}_{n,m+1} + \bar{a}^{\dagger}_{n-1,m} -
a^{\dagger}_{nm} - \bar{a}^{\dagger}_{nm} ) b^{}_{nm} + (a^{\dagger}_{n+1,m} -
\bar{a}^{\dagger}_{n,m-1} + \bar{a}^{\dagger}_{nm} - a^{\dagger}_{nm} )
\bar{b}^{}_{nm} ] \}  \label{eq:1}
\end{eqnarray}
(cf.\ Fig.\ \ref{fig:1}).  Here, the operators
$a^{\dagger}_{nm},\bar{a}^{\dagger}_{nm}$ ($a^{}_{nm},\bar{a}^{}_{nm}$) and
$b^{\dagger}_{nm},\bar{b}^{\dagger}_{nm}$ ($b^{}_{nm},\bar{b}^{}_{nm}$) create
(annihilate) electrons in $s$- and $p$-orbitals, respectively, in the unit cell
labelled $n,m$.  The energies $\epsilon_{\rm s}$ and $\epsilon_{\rm p}$ do not
depend on $n$ and $m$.  Representing the one-electron wave function $\Psi$ as a
linear combination of $s$- and $p$-orbitals,
\begin{equation}
\Psi= \sum_{n,m} (u_{nm} \, a^{\dagger}_{nm} + \bar{u}_{nm} \,
\bar{a}^{\dagger}_{nm} + v_{nm} \,b^{\dagger}_{nm} + \bar{v}_{nm}
\,\bar{b}^{\dagger}_{nm}) |0 \rangle \; , \label{eq:2}
\end{equation}
we write the Schr\"odinger equation $(H-E)\Psi=0$ in terms of a coupled set of
equations for the coefficients $u_{nm}$, $\bar{u}_{nm}$, $v_{nm}$, and
$\bar{v}_{nm}$ as
\begin{eqnarray}
&\ & (\epsilon_{\rm s} - E) u_{nm}+ V (v_{n,m-1} + \bar{v}_{n-1,m} - v_{nm} -
\bar{v}_{nm} ) = 0 \; ,\nonumber \\[0.0cm]
&\ & (\epsilon_{\rm s} - E) \bar{u}_{nm}+ V (v_{n+1,m} - \bar{v}_{n,m+1} +
\bar{v}_{nm} - v_{nm} ) = 0 \; , \nonumber \\[0.0cm]
&\ & (\epsilon_{\rm p} - E) v_{nm}+ V (u_{n,m+1} + \bar{u}_{n-1,m} - u_{nm} -
\bar{u}_{nm} ) = 0 \; , \nonumber \\[0.0cm]
&\ & (\epsilon_{\rm p} - E) \bar{v}_{nm}+ V (u_{n+1,m} - \bar{u}_{n,m-1} +
\bar{u}_{nm} - u_{nm} ) = 0 \; .
\label{eq:3}
\end{eqnarray}
Exploiting the periodicity of the model system, we can find a general solution
of Eqs.\ (\ref{eq:3}) in the form of Bloch sums with coefficients
\begin{equation}
u_{nm} =  u_{00} \, \exp[i (n \phi_{x} + m \phi_{z})] \; , \label{eq:4}
\end{equation}
and similarly for $\bar{u}_{nm}$, $v_{nm}$, and $\bar{v}_{nm}$.  The phases
$\phi_{x}$ and $\phi_{z}$ are related to the wave numbers $k_{x}$ and $k_{z}$
through $\phi_{x} = k_{x} a $ and $\phi_{z} = k_{z} a $. Insertion of Eqs.\
(\ref{eq:4}) into Eqs.\ (\ref{eq:3}) leads to the system of equations
\begin{eqnarray}
\begin{array}{llll}
(\Delta - E) u_{00} - V (Z v_{00} & \!\!\! +  X \bar{v}_{00}&\!\!\!)
= 0 \; ,\\[0.0cm]
(\Delta - E) \bar{u}_{00} + V ( Z^{*} \bar{v}_{00} & \!\!\! - X^{*}
v_{00} &\!\!\!  ) =  0 \; , \\[0.0cm]
(-\Delta - E) v_{00} - V (Z^{*} u_{00} &\!\!\! + X
\bar{u}_{00}&\!\!\!) = 0 \; , \\[0.0cm]
(-\Delta - E) \bar{v}_{00} + V (Z \bar{u}_{00} & \!\!\! - X^{*}
u_{00} &\!\!\! ) = 0 \; ,
\end{array}
\label{eq:5}
\end{eqnarray}
with $Z = 1 - e^{-i \phi_{z}}$ and $X = 1 - e^{-i \phi_{x}}$.  The quantity
$\Delta = (\epsilon_{\rm s} - \epsilon_{\rm p})/2$ is half the energy gap,
and the energy zero has been taken at midgap, i.e., $\epsilon_{\rm s} = -
\epsilon_{\rm p} = \Delta$ (cf.\ left half of Fig.\ \ref{fig:2}).

From Eqs.\ (\ref{eq:5}), the energy dispersion of the homogeneous system is
obtained as
\begin{equation}
E^{2} = \Delta^{2} + 4 V^{2} [\sin^{2}(\phi_{x}/2) + \sin^{2}(\phi_{z}/2)] \; .
\label{eq:6}
\end{equation}
A set of four linearly independent (unnormalized) eigenvectors of Eqs.\
(\ref{eq:5}) is given by
\begin{equation}
u_{00} = 1 \; , \; \bar{u}_{00} = 0 \; , \; v_{00} = - \frac{V Z^{*}}{\Delta
+ E_{\pm}} \; , \; \bar{v}_{00} = - \frac{V X^{*}}{\Delta
+ E_{\pm}} \label{eq:6aa}
\end{equation}
and
\begin{equation}
u_{00} = 0 \; , \; \bar{u}_{00} = 1 \; , \; v_{00} = - \frac{V X}{\Delta
+ E_{\pm}} \; , \; \bar{v}_{00} = \frac{V Z}{\Delta
+ E_{\pm}} \; , \label{eq:6ab}
\end{equation}
where $E_{+}$ $(E_{-})$ is the positive (negative) root of Eq.\ (\ref{eq:6}).

Equation (\ref{eq:6}) describes the dispersion of two adjacent bands in the
two-dimensional tight-binding model for homogeneous semiconductors:  the
positive and negative root correspond to the conduction band and the valence
band, respectively.  In the continuum (long-wavelength) limit $k_{x} a, k_{z} a
\rightarrow 0$, we have
\begin{equation}
E^{2} = \Delta^{2} +  V^{2} a^{2} \, (k_{x}^{2} + k_{z}^{2})
\; . \label{eq:6ac}
\end{equation}
The Dirac-like bulk spectrum of the two-band model is now recovered if the
hopping velocity $w = Va/\hbar$ is identified with the interband velocity
matrix element $v$ \cite{kol98}.  Similarly, the set of eigenvectors
(\ref{eq:6aa}) and (\ref{eq:6ab}) reduces in the continuum limit to the usual
set of spinor solutions of the free Dirac equation (cf., e.g., Ref.\
\cite{sch62}).  The tight-binding model for homogeneous systems (in which
relativistic effects, in particular spin degrees of freedom, are not explicitly
included) thus turns out to be equivalent, in the continuum limit, to the
relativistic free-electron theory.

It is worth mentioning that if the signs of the hopping matrix elements were
chosen such that their sum within a unit cell is zero (cf.\ above), one would
obtain, instead of Eq.\ (\ref{eq:6}), a dispersion of the form
\begin{equation}
E^{2} = \Delta^{2} + 4 V^{2} [\sin(\phi_{x}/2) \pm \sin(\phi_{z}/2)]^{2}
\; . \label{eq:6a}
\end{equation}
In the present context, this dispersion is ruled out owing to its inappropriate
behavior in the continuum limit.  Whether lattice models with dispersions of
the form (\ref{eq:6a}) are useful for other applications appears to be an open
question.

\subsection{One-dimensional heterojunction}

Having established our two-dimensional tight-binding description of homogeneous
semiconductors, we now consider heterojunctions and derive the conditions for
the existence of localized interface states.  Some general features of this
problem can be elucidated already by looking at the {\em one-dimensional}
system represented by a linear chain of atoms oriented perpendicular to the
interface.

For definiteness, we single out the chain formed by the bottom atoms in all
unit cells with $n=0$ in Fig.\ \ref{fig:1}. The unit cells of this chain each
contain two atoms and are labelled by the index $m$ (we omit the index $n=0$).
The Bloch coefficients for the homogeneous one-dimensional system have the form
\begin{equation}
u_{m} = u_{0} \, \exp(i m \phi_{z})\; , \; v_{m} = v_{0} \, \exp(i m \phi_{z})
\; .
\label{eq:7}
\end{equation}
Localized wave functions falling off exponentially on either side of the
heterojunction are characterized by purely imaginary phases in the Bloch
coefficients corresponding to the homogeneous semiconductors 1 and 2.  Writing
\begin{equation}
\phi_{z}^{(1)} = -i \theta_{1} \; , \;  \phi_{z}^{(2)} = i \theta_{2}
\hspace*{0.5cm} (\theta_{1}, \theta_{2} > 0) \; ,
\label{eq:8}
\end{equation}
we obtain from Eqs.\ (\ref{eq:7}) the relations
\begin{eqnarray}
\begin{array}{lll}
u_{m}=u_{-1} \exp[(m+1) \theta_{1}]& , \; \;  v_{m}=v_{-1} \exp[(m+1)
\theta_{1}] \; & {\rm for} \; m < 0 \; ;  \\ u_{m}=u_{0} \exp(-m \theta_{2}) &
, \; \;  v_{m}=v_{0} \exp(-m \theta_{2}) \; & {\rm for} \ m \geq 0 \; .
\end{array}  \label{eq:9}
\end{eqnarray}
In order to connect the decaying wave functions across the interface, we write
down the set of coupled equations for the coefficients $u_{-1}, v_{-1}, u_{0},
v_{0}$ corresponding to the atomic orbitals in the unit cells labelled $m = -1$
and $m = 0$ that sandwich the interface. Using the relations (\ref{eq:9}) to
express the coefficients $v_{-2}$ and $u_{1}$ in terms of the coefficients
$v_{-1}$ and $u_{0}$, respectively, we have from the first and third of Eqs.\
(\ref{eq:3})
\begin{eqnarray}
\begin{array}{llll}
(\Delta_{1} - E) u_{-1} & - \; V_{1} Z_{1 }v_{-1} &  & =
\; 0 \; , \\
(-\Delta_{1} - E) v_{-1} & - \; V_{1} u_{-1} & + \; V' u_{0}  & =
\; 0 \; , \\
(\Delta_{2} + W - E) u_{0} &  - \; V_{2} v_{0} & + \; V' v_{-1}  & = \;
0 \; , \\
(- \Delta_{2} + W - E) v_{0} & - \; V_{2} Z_{2} u_{0} & & =
\; 0 \; ,
\end{array}
\label{eq:10}
\end{eqnarray}
where
\begin{equation}
Z_{j} = 1 - e^{- \theta_{j}}  \; \; ; \; \; j=1,2 \; . \label{eq:10a}
\end{equation}
The quantities $\Delta_{j} = (\epsilon_{\rm s}^{(j)} - \epsilon_{\rm
p}^{(j)})/2$ are half the energy gaps, and $W = (\epsilon_{\rm s}^{(2)} +
\epsilon_{\rm p}^{(2)})/2$ is the work function offset (the energy zero is
taken at midgap in semiconductor 1; cf.\ Fig.\ \ref{fig:2}).

Eliminating in the equations involving the matrix element $V'$ the coefficients
$u_{-1}$ and $v_{0}$ by means of the first and fourth equation and using the
bulk energy dispersions [cf.\ Eq.\ (\ref{eq:6}) specialized to the
one-dimensional case] for the imaginary phases given by Eqs.\ (\ref{eq:8}), we
obtain from Eqs.\ (\ref{eq:10}) a system of two equations for $u_{0}$ and
$v_{-1}$. For its solution, we must require
\begin{equation}
V_{1}^{2} V_{2}^{2} (e^{\theta_{1}} - 1) (e^{\theta_{2}} - 1) + V^{\prime \,
2}(\Delta_{1} - E)(\Delta_{2} - W + E) = 0  \; .
\label{eq:11}
\end{equation}
It is seen from Eq.\ (\ref{eq:11}) that for any choice of the values of the
parameters $V_{1}$, $V_{2}$, and $V'$, there are no solutions within the energy
range $\max\{-\Delta_{1}, -(\Delta_{2} - W)\} < E < \min\{\Delta_{1},
\Delta_{2} + W\}$, i.e., within the fundamental energy gap of the
heterojunction.  We thus find that there are no localized (bound) interface
states in the one-dimensional case, a result that was to be expected since here
no energy can be stored in the motion parallel to the interface.

\subsection{Two-dimensional heterojunction}

We now turn to the {\em two-dimensional} tight-binding treatment of
heterojunctions.  Here, localized states are described by two-dimensional Bloch
waves with real phase $\phi_{x} = k_{x} a$ corresponding to free motion in the
direction parallel to the interface and purely imaginary phases
$\phi_{z}^{(j)}$ given by Eqs.\ (\ref{eq:8}) which correspond to exponential
fall-off in the $z$-direction on either side of the interface.  From Eqs.\
(\ref{eq:4}), we then have
\begin{equation}
u_{nm} =  u_{0,-1} \, \exp[i n \phi_{x} + (m+1) \theta_{1}] \label{eq:12}
\end{equation}
for $m< 0$,
\begin{equation}
u_{nm} =  u_{00} \, \exp[i n \phi_{x} - m \theta_{2}] \label{eq:13}
\end{equation}
for $m \geq 0$, and similarly for $\bar{u}_{nm},$ $v_{nm},$ and $\bar{v}_{nm}$.

As in the one-dimensional case, we now write down the set of coupled equations
for the coefficients corresponding to the atomic sites in the unit cells
labelled $n,m = 0,-1$ and $n,m = 0,0$ (boxes with dashed lines in Fig.\
\ref{fig:1}). Using relations (\ref{eq:12}) and (\ref{eq:13}) to eliminate all
coefficients with labels other than $n,m = 0,-1$ or $n,m = 0,0$, we have from
Eqs.\ (\ref{eq:3})
\begin{eqnarray}
\begin{array}{llllll}
(\Delta_{1} - E) u_{0,-1} & - \; V_{1} (Z_{1} v_{0,-1}  & + \;
X \bar{v}_{0,-1}& \!\!) &= \; 0 \; , \\
(- \Delta_{1} - E) \bar{v}_{0,-1}  & + \; V_{1} (Z_{1} \bar{u}_{0,-1}
 & - \; X^{*} u_{0,-1} &\!\!) & = \; 0 \; , \\
(\Delta_{1} - E) \bar{u}_{0,-1} & + \; V_{1} (\bar{v}_{0,-1} & -
\; X^{*} v_{0,-1} &\!\!)  - \; V' \bar{v}_{00}   &= \; 0 \; , \\
(- \Delta_{1} - E) v_{0,-1} & - \; V_{1} (u_{0,-1}& + \;
X \bar{u}_{0,-1}  &\!\!) + \; V' u_{00}  &= \; 0 \; , \\
(\Delta_{2} + W - E) u_{00} &  - \; V_{2} (v_{00}  & + \; X
\bar{v}_{00} &\!\!) + \; V' v_{0,-1} & = \; 0 \; , \\
(- \Delta_{2} + W - E) \bar{v}_{00} & + \; V_{2} (\bar{u}_{00} &
- \; X^{*} u_{00} &\!\!) - \; V' \bar{u}_{0,-1} &= \; 0 \; , \\
(\Delta_{2} + W - E) \bar{u}_{00} & + \; V_{2} (Z_{2} \bar{v}_{00}  & -
\; X^{*} v_{00} &\!\!) &= \; 0 \; , \\
(- \Delta_{2} + W - E) v_{00} & - \; V_{2} (Z_{2} u_{00}  & + \;
X \bar{u}_{00} &\!\!) &= \; 0 \; . \end{array}
\label{eq:15}
\end{eqnarray}
The coefficients $u_{0,-1}, \bar{v}_{0,-1}, \bar{u}_{00}$, and $v_{00}$ in the
four equations involving the coupling $V'$ can be eliminated with the help of
the remaining equations, leading to a closed system of equations for the
coefficients $\bar{u}_{0,-1}, v_{0,-1}, u_{00}$, and $\bar{v}_{00}$
corresponding to the four atomic sites immediately adjacent to the interface
(box with full lines in Fig.\ \ref{fig:1}):
\begin{eqnarray}
\begin{array}{llllll}
(\Delta_{1} - E) \bar{u}_{0,-1} & - \; V_{1} X^{*} v_{0,-1} & -
\; V' Z_{1} \bar{v}_{00} &= \; 0 \; , \\
(- \Delta_{1} - E) v_{0,-1} & - \; V_{1} X \bar{u}_{0,-1}& + \;
V' Z_{1} u_{00}  &= \; 0 \; , \\
(\Delta_{2} + W - E) u_{00} &  - \; V_{2} X \bar{v}_{00} & + \; V'
Z_{2} v_{0,-1} & = \; 0 \; , \\
(-\Delta_{2} + W - E) \bar{v}_{00} & - \; V_{2} X^{*} u_{00} &
- \; V' Z_{2} \bar{u}_{0,-1} &= \; 0 \; .
\end{array}
\label{eq:16}
\end{eqnarray}
Setting the determinant of this system equal to zero and making use of the
bulk dispersions of the semiconductors 1 and 2,
\begin{eqnarray}
E^{2} &=& \Delta^{2}_{1} + 4 V^{2}_{1} [\sin^{2}(\phi_{x}/2) -
\sinh^{2}(\theta_{1}/2)] \; , \nonumber \\
(E-W)^{2} &=& \Delta^{2}_{2} + 4 V^{2}_{2} [\sin^{2}(\phi_{x}/2) -
\sinh^{2}(\theta_{2}/2)]
\label{eq:16a}
\end{eqnarray}
[cf.\ Eqs.\ (\ref{eq:6}) and (\ref{eq:8})], we obtain the relation
\begin{equation}
V_{1}^{2} V_{2}^{2} (e^{\theta_{1}} - 1)(e^{\theta_{2}} - 1) + V^{\prime
\, 4} (1 - e^{-\theta_{1}}) (1 - e^{-\theta_{2}}) +  2  V^{\prime \, 2} \{
\Delta_{1} \Delta_{2} + 4 V_{1} V_{2} \sin^{2}(\phi_{x}/2) - E(E-W)\} = 0 \; .
\label{eq:17}
\end{equation}
Here, the quantities $\theta_{j}$ can be expressed in terms of the energy $E$
and the phase $\phi_{x} = k_{x} a$ through Eqs.\ (\ref{eq:16a}).  Relation
(\ref{eq:17}) thus determines in an implicit way the in-plane dispersion
$E(k_{x})$ for localized interface states in a semiconductor heterojunction.
It represents a principal result of the present paper.

Up to this point, the matrix element $V'$ describing hopping across the
interface has been considered an independent parameter.  In order to fix the
value of $V'$ within the present model in a consistent way, we relate it to the
bulk matrix elements $V_{1}$ and $V_{2}$ by considering the behavior of the
wave function coefficients $u_{nm}$, $v_{nm}$, $\bar{u}_{nm}$, and
$\bar{v}_{nm}$ when the interface is crossed.

For $u_{nm}$, e.g., this behavior is readily inferred by combining the first
and second of Eqs.\ (\ref{eq:15}), the second of Eqs.\ (\ref{eq:16}), and the
first of the bulk dispersions (\ref{eq:16a}).  With similar reasoning for the
other coefficients, we find
\begin{equation}
\frac{u_{00}}{u_{0,-1}} =
\frac{\bar{v}_{00}}{\bar{v}_{0,-1}} =\frac{V_{1}}{V'} \, e^{\theta_{1}}
\label{eq:17a}
\end{equation}
and
\begin{equation}
\frac{v_{00}}{v_{0,-1}}
= \frac{\bar{u}_{00}}{\bar{u}_{0,-1}} = \frac{V'}{V_{2}} \, e^{-\theta_{2}} \;
. \label{eq:17b}
\end{equation}
Interpreting the set of coefficients referring to a specific unit cell as the
components of a single, four-component wave function, we require all
coefficients to change across the interface by one and the same factor, i.e.,
we equate the right-hand sides of Eqs.\ (\ref{eq:17a}) and (\ref{eq:17b}).
This determines ${V'}^{2}$ as
\begin{equation}
{V'}^{2} = V_{1} V_{2} e^{\theta_{1} + \theta_{2}} \; , \label{eq:17c}
\end{equation}
and all coefficients change by the factor $(V_{1}/V_{2})^{1/2} \,
e^{(\theta_{1} - \theta_{2})/2}$.  In the continuum limit $\theta_{1},
\theta_{2} \rightarrow 0$, the matrix element $V'$ thus appears as the
geometric mean of $V_{1}$ and $V_{2}$.  It is seen below that this choice
for $V'$ is indeed necessary in order to recover the two-band model.

\section{Analysis and comparison with two-band model}

\subsection{Energy spectrum}

\subsubsection{General properties}

Equation (\ref{eq:17}) in conjunction with Eqs.\ (\ref{eq:16a}) reduces
to a quadratic equation for $\sin^{2}(\phi_{x}/2)$ as a function of the
energy $E$ from whose solution one obtains the in-plane dispersion
$E(k_{x})$.  Mere inspection of Eq.\ (\ref{eq:17}) already provides us with
conditions for the existence of solutions for $\theta_{1}, \theta_{2} > 0$,
i.e., conditions for the existence of localized interface states.

It is immediately seen that the spectrum $E(k_{x})$ is bounded by the condition
\begin{equation}
E(k_{x}) > \bar{E}(k_{x})\; , \label{eq:18}
\end{equation}
where $\bar{E}(k_{x})$ is defined by
\begin{equation}
\bar{E} (\bar{E} - W) = \Delta_{1} \Delta_{2} + 4 V_{1} V_{2}  \,
\sin^{2}(\phi_{x}/2) \label{eq:19}
\end{equation}
[we consider here only the {\em positive} (conduction band) branches of the
dispersions for systems {\em without} band inversion, i.e., systems with
$\Delta_{1}, \Delta_{2} > 0$]. The function $\bar{E}(k_{x})$ can be
viewed as a kind of mean in-plane bulk dispersion with energy gap and hopping
matrix element given by the geometric mean of the gaps and matrix elements,
respectively, of the bulk semiconductors 1 and 2, whose in-plane dispersions
$E_{1,2}(k_{x})$ follow from Eqs.\ (\ref{eq:16a}) by setting $\theta_{1} =
\theta_{2} = 0$:
\begin{eqnarray}
E^{2}_{1} &=& \Delta^{2}_{1} + 4 V^{2}_{1} \sin^{2}(\phi_{x}/2)  \; , \nonumber
\\
(E_{2}-W)^{2} &=& \Delta^{2}_{2} + 4 V^{2}_{2} \sin^{2}(\phi_{x}/2) \; .
\label{eq:20}
\end{eqnarray}
Moreover, from Eqs.\ (\ref{eq:16a}) with $\theta_{1}, \theta_{2} > 0$, we have
\begin{equation}
E(k_{x}) < E_{1}(k_{x}) \; ,  \hspace*{0.5cm} E(k_{x})  < E_{2}(k_{x})
\; . \label{eq:21}
\end{equation}
For localized states to occur, it is hence necessary that $E(k_{x})$ lies {\em
below} the bulk dispersions of {\em both} semiconductors.  On the other hand,
condition (\ref{eq:18}) requires $E(k_{x})$ to lie {\em above} the mean
dispersion $\bar{E}(k_{x})$.  Conditions (\ref{eq:18}) and (\ref{eq:21}) can be
simultaneously fulfilled only if the bulk dispersions $E_{1}(k_{x})$ and
$E_{2}(k_{x})$ intersect. The latter condition is equivalent to that of the
two-band model \cite{kol98}.

From Eqs.\ (\ref{eq:20}), it is seen that in the special case where $W = 0$ the
bulk dispersions intersect if and only if the condition
\begin{equation}
0 \leq - \frac{\Delta_{+} \Delta{-}}{4 V_{+} V_{-}}  \leq 1  \label{eq:22}
\end{equation}
holds, where $\Delta_{\pm} = \Delta_{1} \pm \Delta_{2}$, $V_{\pm} = V_{1} \pm
V_{2}$.  This implies, in particular, that $\Delta_{-} / V_{-} < 0$.  If the
bulk dispersions intersect, the mean dispersion lies between the bulk
dispersions for small values of $k_{x}$ (cf.\ Fig.\ \ref{fig:3} below), but
below both bulk dispersions in the vicinity of the point of intersection (thus
the attribute ``mean'' given to the function $\bar{E}$ is not to be interpreted
too narrowly).  At $k_{x} = \pi/a$, where the bulk curves and the mean curve
all attain a maximum, the mean curve lies between the bulk curves if the
conditions
\begin{equation}
- \frac{\Delta_{j} \Delta{-}}{4 V_{j} V_{-}}  \leq 1  \label{eq:22a}
\end{equation}
hold simultaneously for $j=1$ and $j=2$.  In this case, the energy spectrum of
localized interface states, $E(k_{x})$, is restricted to a finite $k_{x}$-range
centered around the point of intersection of the bulk dispersions, with end
points $k_{x}^{(j)}$ $(j=1,2)$ determined by the points of intersection of
$\bar{E}(k_{x})$ with $E_{1}(k_{x})$ and $E_{2}(k_{x})$.  For $W=0$, we have
\begin{equation}
k_{x}^{(j)} =  \frac{2}{a} \, \arcsin ( - \, \Delta_{j} \Delta_{-}/[4V_{j}
V_{-}])^{1/2} \; . \label{eq:23}
\end{equation}
Note that the end points of the spectrum do not depend on the matrix element
$V'$, i.e., on properties characterizing the interface, although the dispersion
determined by Eq.\ (\ref{eq:17}) depends on $V'$.

With $V'$ expressed through Eq.\ (\ref{eq:17c}), Eq.\ (\ref{eq:17}) can be
easily solved in the continuum limit, keeping terms of lowest order in
$\theta_{1},\theta_{2} $ and $\phi_{x}$.  The resulting energy spectrum is
\begin{equation}
E(k_{x}) = \frac{1}{\hbar^{2} k_{x}^{2} w_{-}^{2} + \Delta_{-}^{2}}
\; \{ W (\hbar^{2} k_{x}^{2} w_{1} w_{-} + \Delta_{1}
\Delta_{-}) \pm k_{x} d (\hbar^{2} k_{x}^{2} w_{-}^{2} + \Delta_{-}^{2} -
W^{2})^{1/2} \} \; ,
\label{eq:24}
\end{equation}
where $w_{j} = V_{j} a /\hbar$ are the hopping velocities, $w_{-} = w_{1} -
w_{2}$, and $d = w_{1}\Delta_{2} - w_{2}\Delta_{1}$.  This spectrum agrees with
that of the two-band model (Eq.\ (5) of Ref.\ \cite{kol98}) if the
identification
\begin{equation}
w_{j} \equiv v_{j} \label{eq:25}
\end{equation}
is made, with $v_{j}$ the interband velocity matrix elements of the two-band
model.

The tight-binding model thus generalizes the two-band model and provides a
microscopic foundation for the latter.  In particular, the shallow interface
states identified in Ref.\ \cite{kol98} for the case of non-inverted
semiconductors with intersecting bulk dispersion curves thereby find a firm
theoretical basis.

\subsubsection{Example: GaSb-AlSb}

In order to apply our tight-binding model to specific heterojunctions, we have
to fix the values of the parameters $\Delta_{1}$, $\Delta_{2}$, $W$, $V_{1}$,
and $V_{2}$.  For junctions formed of common III-V materials, the values of the
first three of these parameters can be taken from experiment (cf., e.g., Ref.\
\cite{rho93,par94}).  For the hopping matrix elements $V_{j}$, different sets
of values are available \cite{har89,har99,boy97a,boy97b} either from
theoretical estimates or from fits to specific material properties.  It is not
obvious which one of these sets represents an adequate choice in the context of
the present, simplified model.  We therefore resort here to expressing $V_{j}$
in terms of the velocity matrix elements $v_{j}$ of the two-band model as
$V_{j} = \hbar v_{j}/a$ (cf.\ Sec.\ III.A.1).  The velocity matrix elements, in
turn, are given \cite{kol98} by $v_{j} = (\Delta_{j}/m^{*}_{j})^{1/2}$, where
$m^{*}_{j}$ is the conduction-band effective mass at the $\Gamma$-point.

\begin{figure}
\epsfysize=9.0cm
\epsfbox[-165 60 515 539]{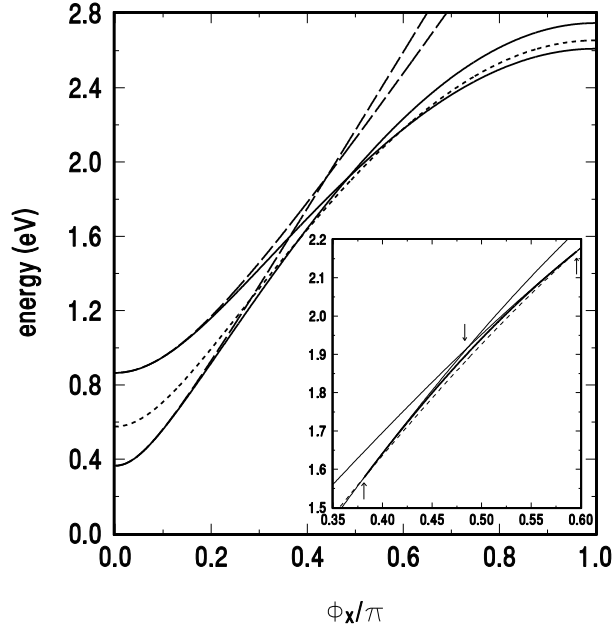}
\caption{Dispersion curves for the conduction band states in the GaSb-AlSb
heterojunction, calculated with the parameter values $\Delta_{1} = 0.365$ eV,
$\Delta_{2} = 0.79$ eV, $W = 0.075$ eV (taken from Table 2 of Ref.\
\protect\cite{rho93}), $V_{1} = 1.3617$ eV, $V_{2} = 1.2041$ eV (derived from
the effective masses given in the second column of Table 6.2 of Ref.\
\protect\cite{har99}; cf.\ text).  Solid curves:  in-plane bulk dispersions
$E_{1}$, $E_{2}$ given by Eqs.\ (\ref{eq:20}).  Short-dashed curve:  mean
in-plane bulk dispersion $\bar{E}$ given by Eq.\ (\ref{eq:19}).  Long-dashed
curves:  Dirac-like in-plane bulk dispersions of the two-band model [cf.\ Eq.\
(\ref{eq:6ac})].  Thick solid curve in the inset:  in-plane energy spectrum of
localized interface states, obtained from the numerical solution of the
tight-binding equations (\ref{eq:16a}) and (\ref{eq:17}) with $V'$ expressed
through Eq.\ (\ref{eq:17c}).  The arrows in the inset mark the intersection of
the bulk dispersions and the intersections of the mean dispersion curve with
the bulk dispersions, respectively.
}
\vspace{0.4cm}
\label{fig:3}
\end{figure}

As a specific example, we consider the conduction band states in the GaSb-AlSb
heterojunction (cf.\ Fig.\ \ref{fig:3}).  With the values for $\Delta_{1}$,
$\Delta_{2}$, and $W$ taken from Ref.\ \cite{rho93} and the (theoretical)
values for $m^{*}_{1}$ and $m^{*}_{2}$ taken from Ref.\ \cite{har99}, we find
that the in-plane bulk dispersion curves $E_{1}$ and $E_{2}$ corresponding to
GaSb and AlSb, respectively, intersect at $\phi_{x} = 1.526$.  The mean
in-plane bulk dispersion curve $\bar{E}$ intersects the curves $E_{1}$ and
$E_{2}$ at $\phi_{x}^{(1)} = 1.200$ and $\phi_{x}^{(2)} = 1.874$, respectively.
Localized interface states exist within the $\phi_{x}$-range between these
limits, with the in-plane energy spectrum shown in the inset of Fig.\
\ref{fig:3}.  The binding energy of these states (as determined by the
difference in energy to the bulk dispersion curves) is of the order of 10 meV
in the vicinity of the point of intersection of the bulk curves.  Since the
Dirac-like dispersion curves of the two-band model intersect at a point close
to the intersection of the tight-binding curves (cf.\ Fig.\ \ref{fig:3}), the
results of the two-band model do not differ much from those of the
tight-binding model.

From considering the example of Fig.\ \ref{fig:3}, it becomes apparent that the
properties of the calculated energy spectrum of interface states depend
sensitively on the values of the model parameters $\Delta_{j}$, $W$, $V_{j}$.
Fairly small changes in one of these values may lead to a large shift in the
position of the point of intersection of the bulk dispersion curves, or may
even remove the intersection.  With regard to the experimental verification of
interface states, it would, therefore, be premature to attach quantitative
significance to the present results.  In order to achieve a fully reliable
description of interface states, refined tight-binding calculations including
the determination of a consistent set of parameters (as outlined, for example,
in Refs.\ \cite{boy96,boy99}) are called for.

\subsection{Wave functions}

In the tight-binding approach, the wave function $\Psi$ represented by the
expansion (\ref{eq:2}) and determined as a solution of Eqs.\ (\ref{eq:16})
describes the electronic motion in the heterojunction as a whole.  The change
of the wave function across the interface is uniquely fixed by the change in
the coefficients expressed by the relations (\ref{eq:17a}) and (\ref{eq:17b})
with $V'$ given by Eq.\ (\ref{eq:17c}).

In the two-band model for abrupt heterojunctions, on the other hand, separate
solutions are obtained for the wave functions in the two bulk regions of the
junction, which are subsequently matched at the interface.  The role of the
matching conditions has been analyzed in Ref.\ \cite{kol98} (cf.\ also Ref.\
\cite{har92}), where conditions for the physical envelope function were
established which relate the discontinuity of the wave function to that of the
interband velocity matrix element across the interface.

In the following, we demonstrate the consistency of the two-band and
tight-binding descriptions with regard to the behavior of the wave functions
across the interface.  We consider the continuum limit of the system of
equations (\ref{eq:5}) describing the homogeneous semiconductor; omitting the
labels attached to the coefficients $u_{00}, \bar{u}_{00}, v_{00},
\bar{v}_{00}$, we introduce linear combinations
\begin{equation}
\psi_{\lambda} = u +i \lambda \bar{u} \; ; \hspace*{0.5cm} \chi_{\lambda} = v +
i \lambda \bar{v} \; , \label{eq:26}
\end{equation}
where $\lambda = \pm 1$.  In the continuum limit, keeping terms up to first
order in $\phi_{z}$ and $\phi_{x}$ and considering $\psi_{\lambda}$ and
$\chi_{\lambda}$ to be continuous functions of $z$ and $x$, i.e.,
$\psi_{\lambda}(z,x), \chi_{\lambda}(z,x) \propto~\exp[i(k_{z}z+k_{x}x)]$, we
replace the wave numbers $k_{z}$ and $k_{x}$ with $-i \partial_{z}$ and $-i
\partial_{x}$, respectively, so that Eqs.\ (\ref{eq:5}) can be expressed in the
form
\begin{eqnarray}
\begin{array}{l} (\Delta - E_{\lambda}) \psi_{\lambda}(z,x) - V a (\partial_{z}
- i \lambda \partial_{x}) \chi_{\lambda}(z,x) = 0 \; , \\[0.3cm]
V a (\partial_{z} + i \lambda \partial_{x}) \psi_{\lambda}(z,x) - (\Delta +
E_{\lambda}) \chi_{\lambda}(z,x) = 0 \; .
\end{array} \label{eq:27}
\end{eqnarray}
These equations are equivalent to the Dirac-like bulk equations of the two-band
model (Eqs.\ (2) of Ref.\ \cite{kol98}) if the hopping velocity $w = V a/\hbar$
is replaced with the interband velocity matrix element $v$ [cf.\ Eq.\
(\ref{eq:6ac})] and if the parameter $\lambda$ is identified with the
eigenvalue of the helicity operator.

Assuming the interface to be located at $z=0$ and using Eqs.\ (\ref{eq:17a}),
(\ref{eq:17b}), and (\ref{eq:17c}), we can now relate the values
$\psi_{\lambda}^{(1)} \equiv \psi_{\lambda}(z = -0)$ and $\chi_{\lambda}^{(1)}
\equiv \chi_{\lambda}(z = -0)$ to the values $\psi_{\lambda}^{(2)} \equiv
\psi_{\lambda}(z = +0)$ and $\chi_{\lambda}^{(2)} \equiv \chi_{\lambda}(z =
+0)$:
\begin{equation}
\frac{\psi_{\lambda}^{(1)}}{\psi_{\lambda}^{(2)}}
=\frac{\chi_{\lambda}^{(1)}}{\chi_{\lambda}^{(2)}} = \left( \frac{V_{2}}{V_{1}}
\right)^{1/2} \; . \label{eq:28}
\end{equation}
These relations are seen to be equivalent to the matching conditions of the
two-band model (Eqs.\ (3) of Ref.\ \cite{kol98}) if we adopt Eq.\ (\ref{eq:25})
to relate the hopping matrix elements $V_{j}$ to the interband velocity matrix
elements $v_{j}$.

\section{Summary and conclusions}

We have presented a tight-binding study of localized interface states in
semiconductor heterojunctions.  Starting from the usual description of
homogeneous III-V semiconductors, we have set up a two-dimensional lattice
model, in which the coupling between the electronic motion perpendicular and
parallel to the interface plane in a junction is explicitly taken into account.
Yet, this model is simple enough in order to allow an essentially analytic
derivation of the conditions for the existence of interface states and a
transparent discussion of their properties.

In constructing the model Hamiltonian, an ambiguity arises in the choice of the
relative sign of the hopping matrix elements in adjacent linear chains of
atoms.  Depending on this choice, two topologically different sytems are
obtained in which the matrix elements within a unit cell add up either to zero
or to a nonzero value.  In the present context, we choose the alternative with
nonzero sum, which leads in the continuum limit to the Dirac-like description
of homogeneous semiconductors within the two-band model.

As a principal result, we have derived an implicit expression for the in-plane
energy dispersion of heterojunctions.  From this expression, conditions for the
existence of localized interface states are inferred, which are equivalent to
those of the two-band model.  In the continuum limit, the energy spectrum of
the two-band model is recovered if the hopping velocities of the tight-binding
model are identified with the interband velocity matrix elements of the
two-band model.  By analyzing the behavior of the tight-binding wave functions
across the interface, we have elucidated the matching conditions for the
physical envelope function in the two-band model.

In conclusion, our tight-binding model for interface states in semiconductor
heterojunctions generalizes the two-band model and provides a microscopic
foundation for it.  Although our model is based on the tight-binding
description of III-V semiconductors, its qualitative features do not appear
too sensitive to specific details of the model and its results are therefore
assumed to be valid for broader classes of materials.  Considering the
possibility of an experimental verification of the type of interface state
identifed in the present work, we have emphasized the need for refined
tight-binding calculations which go beyond the frame of our simple model and in
which consistent parameter sets are used.\\[0.7cm]

\noindent
{\em Note added in proof:} After acceptance of this manuscript, Ref.\
\cite{gor98} was called to our attention. In that paper, interface states in
heterojunctions are treated in a {\em one-dimensional} tight-binding approach
that uses, with respect to the description of III-V semiconductors, more
general assumptions than our two-dimensional model. A relation for the energy
of interface states is obtained that generalizes our relation (\ref{eq:11}) and
for specific parameter values admits solutions within the fundamental energy
gap of the junction. As the electronic motion parallel to the interface is
disregarded in Ref.\ \cite{gor98}, that paper does not deal with the in-plane
dispersion which is the subject of the present work.

\section*{Acknowledgments}
We are indebted to K.\ B.\ Efetov, A.\ Altland, and A.\ P.\ Silin for
useful discussions. One of us (A.\ V.\ K.) gratefully acknowledges support
by  DFG-Sonderforschungsbereich 237 ``Unordnung und gro\ss{}e
Fluktuationen".


\begin{thebibliography}{99}
%
\bibitem{and98}
{\em Mesoscopic Physics and Electronics}, edited by T.\ Ando, Y.\ Arakawa, K.\
Furuya, S.\ Komiyama, and H.\ Nakashima (Springer-Verlag, Berlin, 1998).
%
\bibitem{bas96}
G.\ Bastard, {\em Wave Mechanics Applied to Semiconductor Heterostructures}
(Les \'Editions de Physique, Les Ulis, 1996).
%
\bibitem{fer97}
D.\ K.\ Ferry and S.\ M.\ Goodnick, {\em Transport in Nanostructures}
(Cambridge University Press, Cambridge, 1997).
%
\bibitem{kol98}
A.\ V.\ Kolesnikov, R.\ Lipperheide, A.\ P.\ Silin, and U.\ Wille,
Europhys.\ Lett.\ {\bf 43}, 331 (1998).
%
\bibitem{kol99}
A.\ V.\ Kolesnikov and  A.\ P.\ Silin, Phys.\ Rev.\ B {\bf 59}, 7596
(1999).
%
\bibitem{ivc97}
E.\ L.\ Ivchenko and G.\ E.\ Pikus, {\em Superlattices and Other
Heterostructures} (Springer-Verlag, Berlin, 1997).
%
\bibitem{vol85}
B.\ A.\ Volkov and O.\ A.\ Pankratov, Pis'ma Zh.\ Eksp.\ Teor.\ Fiz.\
{\bf 42}, 145 (1985) [JETP Lett.\ {\bf 42}, 178 (1985)].
%
\bibitem{kor87}
V.\ Korenman and H.\ D.\ Drew, Phys.\ Rev.\ B {\bf 35}, 6446 (1987).
%
\bibitem{pan87}
O.\ A.\ Pankratov, S.\ V.\ Pakhomov, and B.\ A.\ Volkov, Solid State Commun.\
{\bf 61}, 93 (1987).
%
\bibitem{aga88}
D.\ Agassi and V.\ Korenman, Phys.\ Rev.\ B {\bf 37}, 10095 (1988).
%
\bibitem{vol95}
B.\ A.\ Volkov, B.\ G.\ Idlis, and M.\ Sh.\ Usmanov, Usp.\ Fiz.\ Nauk {\bf
165}, 799 (1995)  [Phys.\ Usp.\ {\bf 38}, 761 (1995)].
%
\bibitem{kel63}
L.\ V.\ Keldysh, Zh.\ Eksp.\ Teor.\ Fiz.\ {\bf 45}, 364 (1963) [Sov.\ Phys.\
JETP {\bf 18}, 253 (1964)].
%
\bibitem{wol64}
P.\ A.\ Wolff, J.\ Phys.\ Chem.\ Solids {\bf 25}, 1057 (1964).
%
\bibitem{har89}
W.\ A.\ Harrison, {\em Electronic Structure and the Properties of Solids:
The Physics of the Chemical Bond} (Dover, New York, 1989).
%
\bibitem{har99}
W.\ A.\ Harrison, {\em Elementary Electronic Structure} (World Scientific,
Singapore, 1999).
%
\bibitem{dav92}
S.\ G.\ Davison and M.\ St\c{e}\'{s}licka, {\em  Basic Theory of Surface
States} (Clarendon Press, Oxford, 1992).
%
\bibitem{rho93}
E.\ H.\ Rhoderick, W.\ R.\ Frensley, and M.\ P.\ Shaw, in {\em Handbook on
Semiconductors}, edited by T.\ S.\ Moss (North-Holland, Amsterdam,
1993), Vol.\ 4, p.\ 1.
%
\bibitem{sai78}
G.\ A.\ Sai-Halasz, L.\ Esaki, and W.\ A.\ Harrison, Phys.\ Rev.\ B {\bf
18}, 2812 (1978).
%
\bibitem{aff88}
I.\ Affleck and J.\ B.\ Marston, Phys.\ Rev.\ B {\bf 37}, 3774 (1988).
%
\bibitem{zha90}
F.\ C.\ Zhang, Phys.\ Rev.\ Lett.\ {\bf 64}, 974 (1990).
%
\bibitem{ubb92}
M.\ U.\ Ubbens and P.\ A.\ Lee, Phys.\ Rev.\ B {\bf 46}, 8434 (1992).
%
\bibitem{fra93}
E.\ Fradkin, {\em Field Theories of Condensed Matter Systems} (Addison-Wesley,
Redwood City, 1993).
%
\bibitem{sch62}
S.\ S.\ Schweber, {\em An Introduction to Relativistic Quantum Field Theory}
(Harper \& Row, New York, 1962).
%
\bibitem{par94}
D.\ L.\ Partin and J.\ Heremans, in {\em Handbook on Semiconductors}, edited by
T.\ S.\ Moss (North-Holland, Amsterdam, 1994), Vol.\ 3A, p.\ 369.
%
\bibitem{boy97a}
T.\ B.\ Boykin, G.\ Kliemeck, R.\ C.\ Bowen, and R.\ Lake, Phys.\ Rev.\ B
{\bf 56}, 4102 (1997).
%
\bibitem{boy97b}
T.\ B.\ Boykin, Phys.\ Rev.\ B {\bf 56}, 9613 (1997).
%
\bibitem{boy96}
T.\ B.\ Boykin, Phys.\ Rev.\ B {\bf 54}, 8107 (1996).
%
\bibitem{boy99}
T.\ B.\ Boykin, R.\ K.\ Lake, G.\ Kliemeck, and M.\ Swaminathan, Phys.\ Rev.\ B
{\bf 59}, 7316 (1999).
%
\bibitem{har92}
W.\ A.\ Harrison and A.\ Kozlov, in {\em Proceedings of the 21st International
Conference on the Physics of Semiconductors (Beijing, 1992)}, edited by Ping
Jiang aand Hou-Zhi Zheng (World Scientific, Singapore, 1992), p.\ 341.
%
\bibitem{gor98}
A.\ A.\ Gorbatsevich and I.\ V.\ Tokatly, Pis'ma Zh.\ Eksp.\ Teor.\ Fiz.\ {\bf
67}, 393 (1998) [JETP Letters {\bf 67}, 416 (1998)].

\end{thebibliography}
\end{document}